\documentclass{aa}  

\usepackage{graphicx}
\usepackage{txfonts} 
\usepackage{hyperref}
\usepackage{siunitx}
\DeclareSIUnit{\year}{yr}
\usepackage{booktabs}
\begin{document}

   \title{History-independent tracers}
   \subtitle{Forgetful molecular probes of the physical conditions of the dense interstellar medium}

   \author{J. Holdship
          \inst{1,2}
          \and
          S. Viti\inst{1,2}\fnmsep
          }

    \institute{Leiden Observatory, Leiden University, PO Box 9513, 2300 RA Leiden, The Netherlands\\
              \email{holdship@strw.leidenuniv.nl}
         \and
             Department of Physics and Astronomy, University College London, Gower Street, WC1E 6BT, London, UK
             }

   \date{Received September 15, 1996; accepted March 16, 1997}

% \abstract{}{}{}{}{} 
% 5 {} token are mandatory
 
  \abstract
   {Molecular line emission is a powerful probe of the physical conditions of astrophysical objects but can be complex to model, and it is often unclear which transitions would be the best targets for observers who wish to constrain a given parameter. }
   {We produce a list of molecular species for which the gas history can be ignored, removing a major modelling complexity. We then determine the best of these species to observe when attempting to constrain various physical parameters. }
   {We use  a large set of chemical models with different chemical histories to determine which species have abundances at 1 MYr that are insensitive to the initial conditions. We then use radiative transfer modelling to produce the intensity of every transition of these molecules. We finally compute the mutual information between the physical parameters and all transitions and transition ratios in order to rank their usefulness in determining the value of a given parameter.}
   {We find 48 species that are insensitive to the chemical history of the gas, 23 of which have collisional data available. We produce a ranked list of all the transitions and ratios of these species using their mutual information with various gas properties. We show mutual information is an adequate measure of how well a transition can constrain a physical parameter by recovering known probes and demonstrating that random forest regression models become more accurate predictors when high-scoring features are included. Therefore, this list can be used to select target transitions for observations in order to maximize knowledge about those physical parameters. }
   {}


   \maketitle
%
%-------------------------------------------------------------------

\section{Introduction}
When observing astrophysical environments, molecular line emission offers a wealth of information about the emitting gas; the varying radiative properties of different species means that some are great tracers of specific conditions. For example, gas density can be probed using species that are sensitive to this parameter, such as CS, due to the high critical density of many of their transitions \citep{Bayet2009ExtragalacticSurvey}. On the other hand, gas temperature can  be inferred from species such as H$_2$CO, which gives good estimates of the gas kinetic temperature due to the arrangement of its rotational energy levels \citep{Mangum1993FormaldehydeClouds}.\par
Chemistry complicates and complements this by linking the emission of species to more physical parameters through the abundance of the emitting molecule. Radiative transfer modelling of ions such as C$^+$ can inform an observer of the gas conditions, but these species are most interesting due to the fact that their abundance is tied to the amount of ionizing radiation the gas is exposed to. Considering chemistry can therefore give a different view of which species are important. Species such as CO are useful for their near-constant abundance relative to H$_2$ \citep{Liu2013}, whilst others are considered useful due to the fact that they are only found in specific conditions, such as strong shocks in the case of SiO \citep{jimenez2008,Martin-Pintado1997SiOClouds}.\par

For this reason, observing molecules with the objective of characterizing the emitting gas is common \citep[e.g.][]{Gomez-Ruiz2016,Tanaka2018,Mangum2019}. With measured intensities, one can fit radiative transfer and chemical models to find the underlying physical parameters that give rise to the emission \citep[e.g.][]{Holdship2019SulfurL1157-B1} or one can link measured column densities back to physical parameters using predetermined relationships \citep[e.g.][]{Bovino2020ACores}.\par
However, despite the strong links between the radiative and chemical properties of a species and the observed line intensities, molecules are often chosen opportunistically or based on heuristics from focused chemical modelling studies. Less common is a statistics-based approach to mining a model dataset for useful relationships. For example, \citet{Bron2021TracersGas} use a large set of synthetic data to determine which line ratios are most useful for predicting the ionization fraction of a gas. This approach opens up the possibility of planning observations based on the physical parameters we wish to know with the best possible information on which species will be useful.\par
Whilst this approach is promising, complications can arise when involving chemistry in our modelling of observed line intensities. One particular complication is that the age and history of an object are often unknown. If the age is unknown, then it is often assumed that the abundances have reached steady state, but the history can complicate this. In the best case, the initial abundances can affect how long it takes to reach steady state and therefore whether such an assumption is applicable. In the worst, the initial abundances may affect the steady state value. For example, in hot gas ($\sim$ \SI{200}{\kelvin}), gas phase reactions that both form and destroy methanol are extremely inefficient. Therefore, the gas phase methanol abundance will depend entirely on the initial abundance since the grain surface chemistry is effectively cut off by the high temperature of the gas.\par
The history then is a large concern, and a variety of methods are used by chemical modellers to account for it. In some cases, an evolved object of gas density $n_H$ is modelled by first running a collapse phase where atomic gas is evolved to this density \cite[e.g.][]{Coutens2018} to give reasonable initial abundances. Alternatively, the gas is evolved at the required density in quiescent conditions for a short period, such as 1 Myr, before the science model begins \cite[e.g.][]{Vidal2018,Booth2021AnDisk}. However, neither method offers perfectly realistic starting abundances, and both are therefore a source of uncertainty in any modelling conclusions.\par
As a solution to this, we propose a modelling effort to develop a list of species that we call history-independent tracers (HITs). We define these as molecules that -- under a wide variety of physical conditions -- will quickly converge to the same abundance from a diverse set of initial abundances. That is to say, they `forget' their chemical history in a short time span, making an unknown gas history irrelevant to their modelling. We then go further by taking a statistical approach to determining which physical parameters can be most easily retrieved using the rotational emission of each of these HITs. This will produce a list of species that are relatively simple to model and will allow for strong constraints on specific physical parameters, providing an opportunity for observers to target their observations accordingly.\par
In Sect.~\ref{sec:method} we describe the process followed to generate the list of HITs, including a detailed explanation of the modelling in Sect.~\ref{sec:model}. In Sect.~\ref{sec:results} we present our HITs and the physical parameters for which they are most useful. In Sect.~\ref{sec:discussion} we discuss the limitations of our analysis and assess whether our method produces informative probes. Finally, the work is summarized in Sect.~\ref{sec:conclusions}.

\section{Methods}
\label{sec:method}
The analysis in this work is a four stage process. First,  we repeatedly ran a large grid of chemical models, using different initial abundances each time. Second, we used the output abundances of that grid to find the species that converge to the same abundance within 1 Myr for any given physical conditions regardless of initial abundances: these are the HITs. Then we produced model line intensities of the HITs for every model in our grid. Finally, we took a statistical approach to determining which line intensities are most informative about our underlying physical parameters to determine the use of each HIT. Each of these stages is explained in detailed in the following sections.
\subsection{Chemical modelling}
\label{sec:model}
For this work, we used the time-dependent, gas-grain chemical code UCLCHEM\footnote{\url{https://uclchem.github.io}}\citep{Holdship2017UCLCHEM} to determine the chemical abundances of 304 species as a function of time. We used the most recent release (v2.0) of the code, which includes several major updates. The most important of these is the introduction of three-phase chemistry, in which we treat the surface and the bulk of the ice mantles separately, following \citet{Garrod2011} and \citet{Ruaud2016}.\par
To produce a variety of initial abundances from which our grid of science models will evolve, we ran six different preliminary models to act as the gas `histories'. The final abundances of these histories are used as initial abundances for the science models. In choosing these histories, we aim to make our initial abundances as varied as possible rather than to be strictly realistic, as that would necessitate choosing a type of object to model and would limit the applicability of the analysis.\par
\begin{table}[]
    \centering
        \caption{Histories run for each density in the main model grid. If the initial density is 100.0 cm$^{-3}$, the model collapsed to the density used in the main model. Otherwise, it started at the required density. The final model represents no chemical evolution.}
    \label{table:histories}
    \begin{tabular}{ccc}
    \toprule
    Initial Density / cm$^{-3}$ & Temperature / K & Age\\
    \midrule
    100.0 & 10.0 & Freefall time\\
    100.0 & 30.0 & Freefall time\\
    100.0 & 10.0 & Freefall time + 1 Myr\\
    Final & 10.0 & 1 Myr\\
    Final & 30.0 & 1 Myr\\
    - & - & 0 Myr\\
    \bottomrule
    \end{tabular}
\end{table}
We chose to mimic commonly used chemical modelling approaches by using collapse models and static clouds for our histories. Additionally, we included the extreme case of initializing the science models with purely atomic gas. This results in preliminary models where complex chemistry and ice phase abundances have built up to different degrees and represent a wide span of possible initial abundances. The histories are given in Table~\ref{table:histories}. Models with an initial density of \SI{100}{\per\centi\metre\cubed} increase in density following a simple free-fall equation \citep{Rawlings1992} up to the density of the science model. The other models are initialized at the required density and simply evolve under constant gas conditions.\par
\begin{table}[]
    \centering
        \caption{Physical parameters varied across the grid of models with their ranges and whether linear or log space sampling was used.}
    \label{table:grid}
    \begin{tabular}{ccccc}
    \toprule
    Variable& Unit  & Range & Sampling\\
    \midrule
    Density & \si{\per\centi\metre\cubed} & \num{e4}-\num{e7} & log\\
    Temperature & K  &  10--300 & linear\\
    F$_{UV}$ & Habing & 1--\num{e3} & log \\
    A$_V$ & $mag$ & 1--10 & log \\
    $\zeta$ & \SI{1.3e-17}{\per\second} & 1--\num{e3} & log \\
    \bottomrule
    \end{tabular}

\end{table}
These histories were then used to provide the starting abundances for a grid of models in which we varied the local UV field, the cosmic ray ionization rate, gas temperature, and gas density. We ran each model for \SI{1}{\mega\year} and retrieved the final abundances for all species. The parameter ranges are given in Table~\ref{table:grid}. Since the histories must be calculated for each gas density, we used four fixed density values. For the other parameters, we sampled 1000 points from the parameter space using Latin hypercube sampling. This results in a grid of 4,000 models, which we ran for each of the six histories.\par
\subsection{Finding HITs }
\label{sec:hit_finding}
With the abundances after \SI{1}{\mega\year} of chemical evolution from diverse histories, we searched for species that give the same abundance regardless of their history. To do this, we first grouped our database by species and physical parameter values and calculated the standard deviation of the log abundances. This standard deviation is therefore the average amount the abundance of a species varies across our histories for a given set of physical parameters. An ideal HIT will have a standard deviation of zero because the same parameters will give the same abundance regardless of history.\par
We then took all these standard deviations and calculated the mean across all physical conditions for a given species. This mean will be zero if a species' variation due to history is zero for all physical conditions but will increase if a species is sensitive to its history under certain conditions.\par
A perfect HIT will have a mean deviation of 0.0, but we will also consider the percentage of models in our grid for which a species has a deviation of $< 0.5~dex$. This less stringent requirement will help identify species that are typically history independent in the event no perfect HITs are found.\par
We imposed three additional constraints when selecting HITs. Firstly, we considered gas phase species only. Whilst grain surface chemistry is included in our model, our goal is to recommend observational targets from which one can infer gas properties. Species in the ices are not easy to detect and are complex to model. Secondly, for a species to be considered a HIT, it must have a median fractional abundance $>$\num{e-12} so that it is likely to be observable. Finally, collisional data must be available in the LAMDA database \citep{Schoier2005LAMDA}.
\subsection{Model intensities}
Our goal is ultimately to link observables to underlying physical parameters using chemistry that we can demonstrate is simple to model. Therefore, we used our grid of models to produce line intensities for every transition of the HITs for which collisional data are available in the LAMBDA database \citep{Schoier2005LAMDA}. As a simplifying choice, we used the ALMA frequency range to limit the range of transitions we consider.\par
We produced the line intensities using SpectralRadex\footnote{\url{https://spectralradex.readthedocs.io}} \citep{Holdship2021TheALCHEMI,VanderTak2007}. Our RADEX inputs are the H$_2$ density, the gas temperature, and a species column density calculated by multiplying the radius of the model cloud by its density and the species abundance. History-independent tracers can still vary slightly in abundance for a given set of physical parameters over the different histories in our grid, so we took the average for our RADEX input. In addition to these model-specific parameters, we chose an arbitrary line full width at half maximum of \SI{5}{\kilo\metre\per\second} and an ortho-to-para ratio of 3 when densities of the isotopomers of H$_2$ are required for RADEX. In many conditions, the H$_2$ ortho-para ratio will differ from this value. We therefore computed the line intensities with a ratio of 0.1 to determine whether a change in this value would affect our results, and we found no change to our conclusions, indicating that this analysis is not sensitive to this assumption.\par
In order to make our line intensities more realistic, we added random noise to every line flux. Without noise, a line that would be undetectable in most cases may still be found to be important, and so we simply needed to add sufficient noise to make weak lines uninformative. We assumed an rms value of 0.05 K and a velocity resolution of \SI{0.5}{\kilo\metre\per\second} to calculate the uncertainty on an integrated emission taken over 20 channels. This is $\sim$ \SI{1}{\kelvin\kilo\metre\per\second}. We consider this to be a reasonable value for the noise level on galactic observations,  and it is certainly sufficient to mask any information contained in lines much lower in intensity than this value.\par
Finally, we accounted for the fact that low resolution or large line widths can result in multiple transitions being blended. We combined the flux of any transitions of a given species that are within 10 MHz of one another. This typically has the effect of combining hyperfine lines into a single flux. The result of this whole process is a table of physical conditions and the corresponding intensities of every transition of the HITs.
\subsection{Connecting HITs to physical parameters}
The final step in the process was to determine the use of each HIT by finding which HITs are most informative about each physical parameter in our models. To do this, we estimated the mutual information between each transition's fluxes and the target variables. Mutual information is a measure of how much information one variables provides about another \citep{Cover1991EntropyInformation}, and ranking variables by the mutual information is a common feature selection algorithm that has been shown to be effective in minimizing prediction error \citep{Frenay2013IsRegression}. \par
Before evaluating the mutual information, we scaled the line intensities using the QuantileTransformer class from the scikit-learn python library \citep{scikit-learn}. This assigns every intensity a new value from 0-1 depending on the percentile it belongs to when compared to all intensities of the same transition. This has the benefit of being robust to outliers, which is important as RADEX will occasionally predict spurious, near-infinite values that are compressed to the range 0.99-1.0.\par
In addition to the fluxes of every transition of our HITs, we calculated all possible line ratios from the transitions because line ratios are commonly used as probes of physical parameters and are likely to be informative \citep[e.g.][]{Green2016DenseGalaxies,Viti2017MolecularEnvironments,Hacar2020HCN-to-HNCISM}. We included all possible combinations because we find that the best line ratios are not necessarily the ones that combine the most informative single transitions, and therefore, the importance of line combinations must be calculated separately to find the most useful features. Henceforth, we refer to individual ratios and transitions as features. \par
We then used the mutual\_info\_regression function from scikit-learn between all  features and physical variable values. The result is a table of every pair of species transition and target physical parameters with their mutual information. From this, the best transitions to observe in order to constrain a given parameter can be found and the average information per transition can be calculated to determine which species are typically most useful for measuring a given physical parameter.\par
\section{Results}
\label{sec:results}
\subsection{Identifying HITs}
\label{sec:hitlist}
Following the procedure described in Sect.~\ref{sec:hit_finding}, we find 52 gas phase species that have a mean standard deviation across all physical conditions that is less than 0.5 dex. However, there exists at least one set of physical conditions for every species where the standard deviation between histories is greater than 0.5 dex. This means there are no perfect HITs. Instead, we look at the percentage of parameter combinations for which a species abundance deviates by less than 0.5 dex between histories.\par
We find 43 gas phase species that have a standard deviation of $< 0.5$ dex in at least 80\% of cases. A further 30 species fit this criterion but have a median abundance of $<$\num{e-12} and so are discarded as they will typically be too low in abundance to be observed easily. Of the 43 acceptable HITs, we present in Table~\ref{table:hitlist} the 23 species for which collisional data are available in the LAMDA database \citep{Schoier2005LAMDA}.\par
\begin{figure*}
    \centering
    \includegraphics[width=0.9\textwidth]{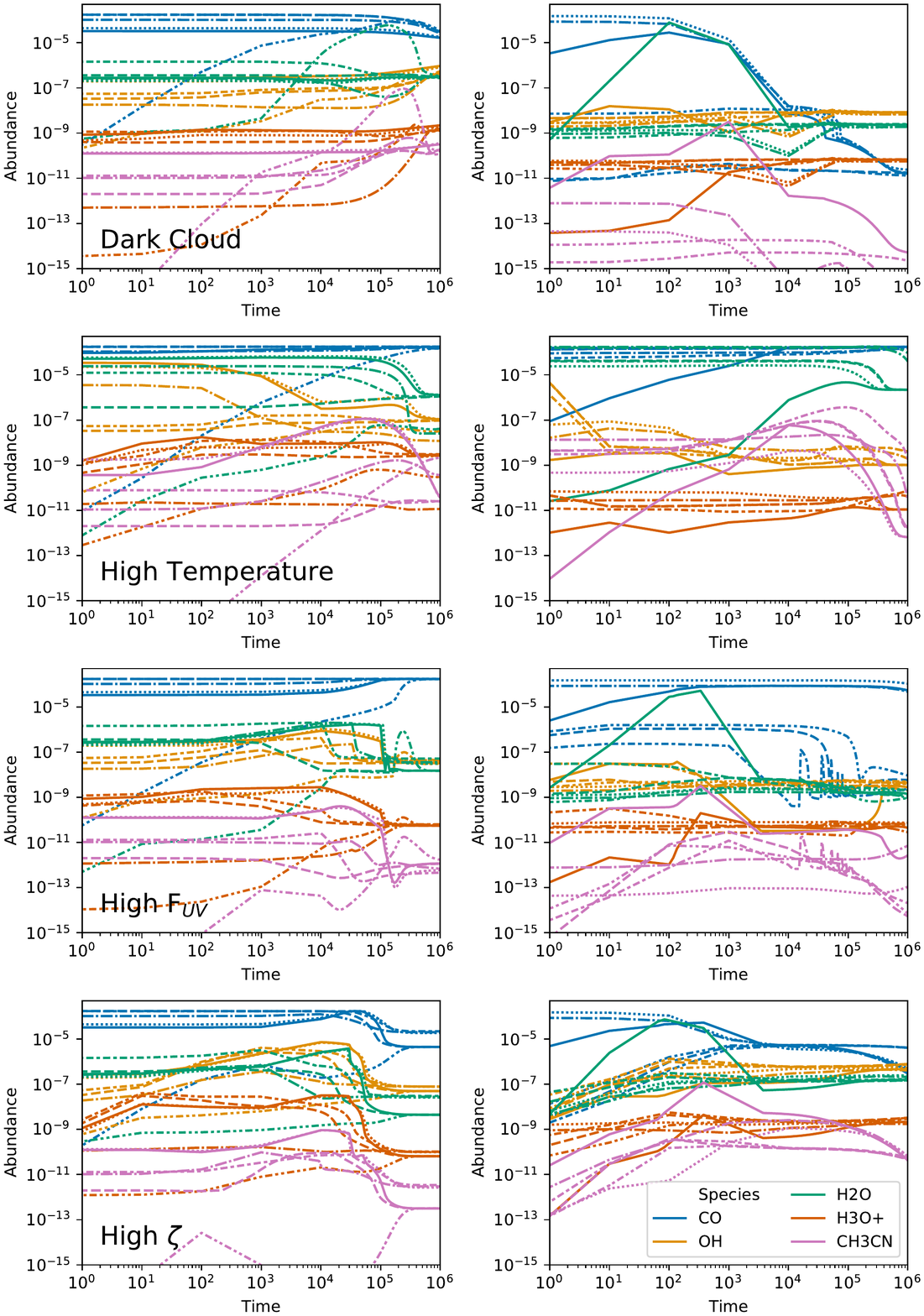}
    \caption{Fractional abundance, with respect to the total number of hydrogen nuclei, of selected HITs as a function of time. Different line styles represent different initial abundances due to different histories. The colour indicates the species.}
    \label{fig:abundances}
\end{figure*}

\begin{table}[]
    \centering
    \caption{Species that have an abundance after \SI{1}{\mega\year} that varies by less than 0.5 dex depending on initial conditions in at least 80\% of cases. }
\label{table:hitlist}
\begin{tabular}{lll}
\toprule
    Species & Percentage &          Convergence Time / yr \\
\midrule
         CO &         97 &  \num{1.0e+03} \\
         OH &         96 &  \num{1.0e+04} \\
        HCL &         92 &  \num{1.0e-01} \\
         CS &         91 &  \num{1.0e+04} \\
         CH &         89 &  \num{9.0e+04} \\
        HCN &         88 &  \num{5.1e+04} \\
         CN &         88 &  \num{1.7e+04} \\
      C$^+$ &         88 &  \num{4.1e+04} \\
        HNC &         88 &  \num{7.6e+04} \\
     H$_2$O &         87 &  \num{3.3e+04} \\
 H$_3$O$^+$ &         87 &  \num{1.0e+04} \\
        HCO &         86 &  \num{1.1e+05} \\
         NO &         85 &  \num{4.4e+04} \\
    HCO$^+$ &         85 &  \num{1.5e+04} \\
 N$_2$H$^+$ &         85 &  \num{2.8e+04} \\
     NH$_3$ &         84 &  \num{7.8e+04} \\
   CH$_3$CN &         84 &  \num{1.0e+05} \\
     H$_2$S &         83 &  \num{8.6e+04} \\
        SIO &         83 &  \num{1.0e+04} \\
     CH$_2$ &         83 &  \num{1.0e+05} \\
    HCS$^+$ &         82 &  \num{1.0e+04} \\
         SO &         81 &  \num{8.0e+04} \\
    H$_2$CO &         81 &  \num{1.0e+05} \\
\bottomrule
\end{tabular}
\end{table}
In Fig.~\ref{fig:abundances} we show the time evolution of a selection of the HITs for various physical parameters. The left column shows models at a density of \SI{e4}{\per\centi\metre\cubed}, and the right column shows similar models at \SI{e7}{\per\centi\metre\cubed}. The top row shows a model with all parameters at the low end of the ranges given in Table~\ref{table:grid}, and each subsequent row pushes one parameter to the highest end of the range. Each line style in the plots represents a model run with the same physical parameters but from different initial abundances. These traces start with differences of orders of magnitude but almost always converge before 1 MYr, at which point we can claim the species has forgotten its chemical history. Typically, the HITs take less time to converge in high density models. \par
The time taken for a HIT to forget its history is an important result, and we call it the convergence time. If the convergence time is shorter than the timescale over which an object's chemistry is to be modelled, the history of the gas need not be considered. The median time taken for each HIT's abundance to converge is given in Table~\ref{table:hitlist}. The smaller this number, the more quickly the species tends to forget its history. 
\subsection{Tracers}
In this section we discuss each physical parameter in turn, listing the best probes of that parameter and discussing the potential benefits and drawbacks of those probes. However, we additionally provide the full table of transitions, parameters, and their mutual information as supplementary material so that observers can select their own targets based on the mutual information and guidance provided in Sect.~\ref{sec:selecting}. Our results are also searchable at our website.\footnote{\url{https://uclchem.github.io/hits}}\par
\subsubsection{Temperature}
\label{sec:temperature}
The most useful features for estimating the temperature are almost entirely line ratios. This is to be expected as the relative excitation of lines with different properties is strongly dependent on the temperature. However, what may be more unusual is that line ratios between transitions of different species become very important when chemistry is included as the relative abundances become informative rather than being free parameters.\par
Ratios between various transitions of CO and many of transitions of CH$_2$, CH$_3$CN, H$_2$S, and NO all have mutual information scores between 1.25 and 1.3, making them dominate the ranked list of most informative features. In fact, CO is easily the most informative species as not only is it part of every single ratio in the top 100 most informative features, its transitions above \SI{300}{\giga\hertz} are similarly informative. \par
It is worth considering whether our method assigns high importance to known temperature probes. The HCN-to-HNC ratio has recently been proposed as a temperature probe \citep{Hacar2020HCN-to-HNCISM}, and we find that ratios of these species have mutual information scores in the range 0.8-0.95. The use of ratios between transitions of H$_2$CO is also an established temperature probe \citep{Mangum1993FormaldehydeClouds}, and we find a mutual information score of 0.6 between many of those line ratios and the temperature. These high mutual information scores show that our method is assigning larger importance to useful features. However, our combination of chemistry and radiative transfer allows us to find even more sensitive probes as many of our ratios have higher importance. This is reflected in Sect.~\ref{sec:recovery}, where we show the uncertainty in temperature predictions with our most useful features, finding that they are significantly smaller than those obtained from H$_2$CO ratios in \citet{Mangum1993FormaldehydeClouds}. 
\subsubsection{Gas density}
Gas density is also best probed through ratios.\ However, at 0.85, the top ranked feature has a mutual information score significantly lower than that of the gas temperature, indicating that constraining this parameter will require more transitions than the temperature.\par
Similarly to the temperature, the list of most important features is dominated by a single species, SO in this case. Ratios among SO transitions and between SO and NO, CH$_2$, N$_2$H$^+$, and NO are all highly ranked. However, one notable difference from the case of temperature is that the usefulness of these ratios appears to be independent of inversion. All useful CO ratios for determining temperature use a CO line as the numerator, but ratios with SO for the gas density appear to have similar mutual information regardless of orientation.\par
SO also has several transitions -- at 66.03, 428.11, and \SI{384.53}{\giga\hertz} -- that are independently informative of the density.  These are useful for observations with limited bandwidth, particularly the \SI{66.03}{\giga\hertz} transitions, which can be combined with a CH$_2$ transition at \SI{69.04}{\giga\hertz}.\par
Comparing this to known density probes, we find that all of the most informative single transitions belong to the dense gas tracer NH$_3$ \citep{Tafalla2004}. Furthermore, excluding SO, which appears to be a new probe of density, we find that combinations of CS, NH$_3$, and N$_2$H$^+$ are all highly ranked and are known dense gas tracers \citep{Bayet2009ExtragalacticSurvey,Punanova2018AstronomyFilament,Johnstone2010DENSECORES}.
\subsubsection{Visual extinction}
Unsurprisingly, given its known photodissociation in regions of low visual extinction, CO is the most informative species for predicting $A_V$. Many ratios with H$_2$CO, H$_2$S, H$_3$O$^+$, and HCO$^+$ have information scores between 0.9 and 1.0. Each of these ratios makes use of a CO transition as the numerator.\par
Unlike the previous two parameters, no individual transition is particularly informative for the visual extinction. The CO transition at \SI{230.54}{\giga\hertz} has a mutual information of just 0.64, approximately two-thirds of the value of the best ratios. There are also many NO transitions that have a similar information score as the CO transitions; however, NO does not appear in any of the most useful ratios. This is an important note: simply combining informative features does not necessarily produce useful features, and features with individually low information scores often combine to make something very informative.\par
\subsubsection{Cosmic ray ionization rate}
The cosmic ray ionization rate has one of the lowest mutual information scores with its most useful feature. Ratios of CO/H$_3$O$^+$, H$_2$O/H$_3$O$^+$, CO/HCO$^+$, and HCN/HNC have mutual information scores between 0.7 and 0.8. The HCO$^+$ and H$_3$O$^+$ lines are also the most individually informative transitions, likely due to the sensitivity of their abundances to the ionization rate. With that in mind, the most useful ratios appear to be the ratio of these ionization-sensitive species to something ubiquitous and less sensitive, such as CO and H$_2$O.\par
Whilst these ratios are the most useful, there are many H$_2$S/CO ratios that are only slightly less informative. Almost all of these ratios use the CO \SI{230.54}{\giga\hertz} transition as the denominator. This makes the \SI{228.56}{\giga\hertz} H$_2$S line particularly useful as a small observing bandwidth would capture both lines and a slightly larger window may also capture an additional H$_2$ transition at \SI{204.14}{\giga\hertz}. \par
Molecular tracers of the cosmic ray ionization rate are less common in the literature. However, \citet{Izumi2016SUBMILLIMETER-HCNGALAXIES} find that the HCN-to-HCO$^+$ ratio is a good indicator of active galactic nucleus activity based on its response to X-ray irradiation. Since, chemically, there is little difference between the effects of  cosmic ray and X-ray radiation \citep{Viti2014MolecularGas}, it is likely this ratio would be useful for the cosmic ray ionization rate. We find that ratios of these species have mutual information scores between 0.5 and 0.6, putting them in the top percentile of features.\par
\subsubsection{External UV flux}
The external UV field is the physical parameter that has the least mutual information with the features. The best features have a mutual information score of 0.74. Surprisingly, given how strongly the two parameters interact in UCLCHEM, the mutual information between each feature and the UV field does not correlate with the information between the transitions and the visual extinction. \par
The most useful features for estimating the UV field are ratios of HCN/HNC and their inverse, CO/H$_3$O+, CO/HCO$^+$, and H$_2$O/H$_3$O+. We also find a remarkably similar situation to the cosmic ray ionization rate, in the sense that many ratios of H$_2$S/CO are informative about the UV field.\par
However, we note that none of these are as informative as the top species listed for other variables and that UCLCHEM is not designed to properly treat photon-dominated regions. Thus, if the UV field is of particular interest, this is unlikely to be the best modelling approach and a PDR model should be used instead.
\section{Discussion}
\label{sec:discussion}
\subsection{Multipurpose HITs}
Regarding the individual lists of the most informative features for each physical parameter, many features appear in multiple lists. Thus, we investigated which species and ratios of species' transitions provide the most information on the most physical parameters. To do this, we ranked each species by its mean mutual information with a physical parameter across all of its transitions and each pair of species by the mean of their mutual information across all of their possible transition ratios. \par
\begin{table*}[]
    \centering
        \caption{Five most informative species pairs for each physical parameter. An X indicates that a species is in the top five for a given parameter.}
    \label{tab:top_n}
\begin{tabular}{lllllll}
\toprule
         Molecules & n$_{H2}$ / cm$^{-3}$ & Temperature / K & $\zeta/\zeta_0$ & F$_{UV}$ / Habing & A$_V$ / mag \\
\midrule
                 CO-H$_2$S &                      &               X &               X &                 X &           X \\
     CO-H$_3$O$^+$ &                      &                 &               X &                 X &           X \\
             CO-NO &                      &                 &               X &                 X &             \\
           HCN-HCN &                      &                 &               X &                 X &             \\
         CH$_2$-CO &                      &               X &                 &                   &             \\
         NH$_3$-NO &                    X &                 &                 &                   &             \\
H$_3$O$^+$-HCO$^+$ &                      &                 &                 &                 X &             \\
    H$_3$O$^+$-HCN &                      &                 &                 &                   &           X \\
         H$_2$O-NO &                    X &                 &                 &                   &             \\
 H$_2$O-H$_3$O$^+$ &                      &                 &               X &                   &             \\
        CO-HCS$^+$ &                      &               X &                 &                   &             \\
 CH$_2$-H$_3$O$^+$ &                    X &                 &                 &                   &             \\
        CO-HCO$^+$ &                      &                 &                 &                   &           X \\
        CO-H$_2$CO &                      &                 &                 &                   &           X \\
                CO &                      &               X &                 &                   &             \\
       CH$_3$CN-CO &                      &               X &                 &                   &             \\
     CH$_2$-NH$_3$ &                    X &                 &                 &                   &             \\
             NO-NO &                    X &                 &                 &                   &             \\
\bottomrule
\end{tabular}

\end{table*}
In every case, species ratios were the most informative, and so we present the top five species pairs for each physical parameter in Table~\ref{tab:top_n}. An X indicates that the ratios of a given pair of species are in the top five most informative ratios for a physical parameter.\par
We find that ratios of CO and H$_2$S are particularly useful. Transition ratios from these species have high mutual information with four physical parameters: the gas temperature, the cosmic ray ionization rate, the UV field, and visual extinction. CO is an extremely useful species in general, appearing many times in different pairs throughout the table.\par
Another interesting species is SO, which appears in every species pair in the top five list for gas density, including in ratios between its own transitions. This is convenient as it means that focusing on transitions of SO could be an effective way to constrain the gas density. However, the lack of a crossover between the top five list for gas density and the other parameters does make it difficult to constrain all parameters at once.
\subsection{HIT accuracy}
\label{sec:recovery}
 In previous sections we have listed the  most informative features for predicting a given parameter. However, whilst one expects a transition with a high mutual information score to be strongly predictive of the parameter, it is not clear how our predictive performance varies with mutual information scores. To investigate how accurately these parameters can be recovered, we trained predictive models for each parameter and evaluated their performance.\par
To properly evaluate our features, we required a statistical model that is likely to perform well if the inputs it is given are predictive of a given parameter. For this reason, we chose random forest regressors, which are flexible models that typically perform well even on complex tasks. For each parameter, we trained a random forest on 75\% of our dataset and tested it on the rest using one, five, and ten features selected from the mutual information table as inputs. Those features were chosen by prioritizing the features that share the most information with the target variable but only allowing one ratio per pair of species to be included (see Sect.~\ref{sec:selecting}). \par
\begin{figure*}
    \centering
    \includegraphics[width=\textwidth]{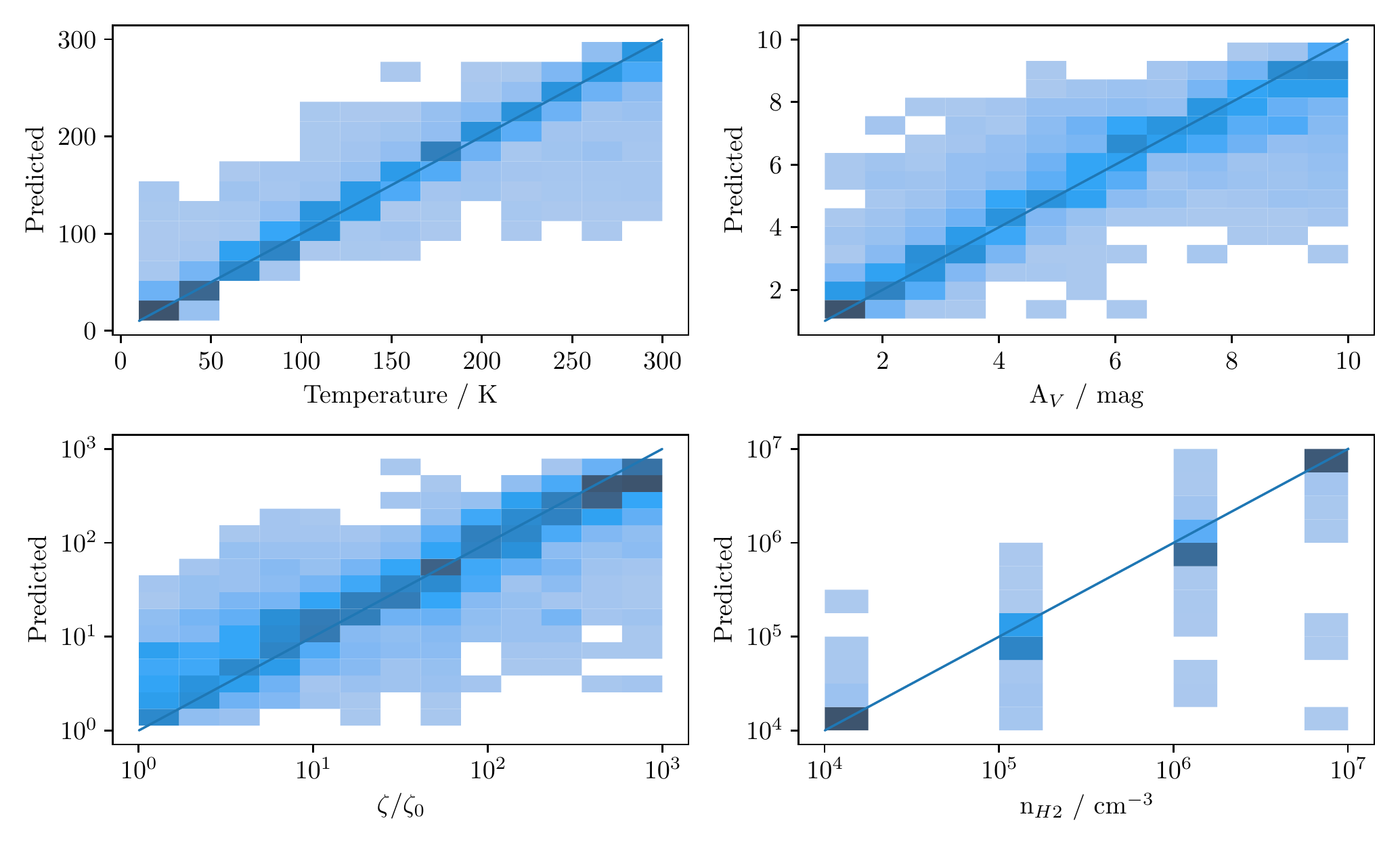}
    \caption{Histograms of the predicted against real value of each parameter from the model dataset. Darker areas show more frequent combinations, and the diagonal line in each plot shows a perfect prediction. The density is sparsely sampled, hence the discrete x axis. The cosmic ray ionization rate is given in units of $\zeta_0$ = \SI{1.3e-17}{\per\second}.}
    \label{fig:predictions}
\end{figure*}

\begin{table}[]
    \centering
        \caption{Mean absolute percentage error when predicting each parameter from its top one, five, and ten features.}
    \label{table:accuracy}
\begin{tabular}{lrrr}
\toprule
            Parameter & \multicolumn{3}{c}{Mean Error} \\
                      &          1 &       5 &      10 \\
\midrule
     Temperature / K &      75.4\% &  14.8\% &   7.0\% \\
         A$_V$ / mag &      30.9\% &  20.9\% &  16.8\% \\
n$_{H2}$ / cm$^{-3}$ &     223.3\% &  16.0\% &  18.3\% \\
     $\zeta/\zeta_0$ &     521.9\% & 118.5\% &  85.4\% \\
              R / pc &     469.2\% & 144.3\% & 103.9\% \\
   F$_{UV}$ / Habing &    1200.1\% & 562.8\% & 489.8\% \\
\bottomrule
\end{tabular}

\end{table}
We present the mean absolute percentage error of the random forest model when trained with each subset of features in Table~\ref{table:accuracy}. We also show the performance of the models on the test data in Fig.~\ref{fig:predictions}, which uses a colour scale to indicate how often a prediction takes a particular value for a given real value when the random forests trained on the top five transitions are used to make predictions. A perfect model would show a dark shaded region along y=x, and in most cases our predictors follow that trend very closely.\par
In the cases of gas temperature, density, and visual extinction, even the top five features produce good estimates. This can be seen in Fig.~\ref{fig:predictions} from the fact that the y=x line has many models, as indicated by the dark shading. The cosmic ray ionization rate on the other hand is less well constrained, but the uncertainty should be considered in the context of astronomical observations as an 85\% error means knowing the ionization rate to within a factor of two to ten, depending on whether the model over- or underestimates.\par
However, even considering the large errors that would still be useful in astronomical applications, the UV field is not well captured at all. In fact, the UV field is not plotted in Fig.~\ref{fig:predictions} because the random forest typically predicts the mean value across all models in the dataset, showing no trend between the predictions and the value. This is a strategy that machine learning models often choose when nothing useful can be learned from the data. The UV field is the parameter with the lowest mutual information with the features, so this should be expected. In fact, Table~\ref{table:accuracy} is sorted by the error when using ten features but happens to be in order of descending mutual information, showing that it is a useful metric for choosing features.\par 
One interesting aspect of this test is that extending it to the top ten features produces lower errors in most cases, but the model actually performs slightly worse in the case of density. This is due to an overlap in the information contained in the first five and second five features for this parameter. As a result, the noise increases with additional features but the information does not, causing the model to perform worse. This highlights the problem with using regression models as a method of feature selection. The behaviour of the random forest algorithm has produced an artificial drop in fit quality that could lead to incorrect conclusions about which transitions to measure. In a standard parameter inference, the uncertainty would not increase, and so the random forest error is misleading. \par
An extension of this is that whilst a low error indicates that the features contain enough information to constrain the parameters, they should not be taken as the best possible measurements one could obtain with the features. Even the best statistical models will perform worse than an appropriate physical model, and thus fitting measured ratios with UCLCHEM and RADEX is likely to produce smaller uncertainties than the random forest predictions. Furthermore, these errors assume no knowledge of the other physical parameters, beyond what is implicitly contained in the transitions used for the model. However, there are strong degeneracies between the physical parameters in these models, which must inflate the uncertainty. Thus, if previous work or alternative measurements already constrain one or more physical parameters (as if often the case), measurements of a HIT's transitions would likely give better constraints on their respective parameters than implied by the errors given here. For example, a density measurement using SO transitions would likely benefit from constraints on the temperature provided by previous CO measurements.\par
\subsection{Selecting features}
\label{sec:selecting}
The goal of this work was to find species that are simple to model and will be effective in determining various physical parameters. Thus, we present a list of transitions and ratios with their mutual information scores, having shown that this is an adequate measure of their usefulness. However, there are some caveats to bear in mind when selecting targets for observation.\par
Firstly, features share mutual information not only with the target parameters but also with each other. This is the reason the models in Sect.~\ref{sec:recovery} use only one ratio per pair of species, and typically we find extremely high mutual information and Spearman correlation coefficients greater than 0.9 between different ratios of the same pair of species. Thus, adding multiple ratios from the same pair does not increase the information you have on the target parameter. Whilst it is beyond the scope of this work to compute the mutual information of all pairs of features, we suggest that observers favour either transitions from unique species or transitions with very different excitation conditions when using the mutual information table to select target transitions.\par
Secondly, one of our findings is that many ratios of informative transitions carry much less mutual information than ratios of relatively uninformative transitions. This finding indicates that if more than one transition or ratio is to be observed, the best combination is not obvious. However, the results of our random forests indicate that combinations of the most individually informative features do make for effective probes of a physical parameter even if they are not the best possible combination. \par
Finally, we have taken no mitigating steps to ensure that all the ratios in our feature list are mutually observable across a wide range of transitions. We argue that this should not be a problem on two grounds. First, the mutual information score will be small if one transition is approximately zero across a large area of parameter space. When one transition is most often below the noise level, the ratio will be very large or very small across a large area of parameter space, and therefore it would have low mutual information with the physical parameters. Thus, any highly ranked ratio must vary strongly. Second, the noise we add to all intensities before calculating the mutual information would remove any feature that typically has a small enough flux to be noise dominated. Therefore, it is likely that any highly ranked ratio is between transitions that are often mutually excited. Moreover, one transition being below the noise level should be informative in itself.\par
\subsection{Limitations of the models}
In this work we use a modelling approach that assumes an object can be modelled using one or more single-point chemical models, each coupled to a simple radiative transfer or local thermodynamic equilibrium model to produce the observed emission. Despite being highly simplified, these assumptions are commonly used to model objects as diverse as molecular clouds \citep{Vidal2018}, pre-stellar cores \citep{Belloche2016,Vasyunin2017,Jin2020FormationChemistry}, protostellar disks \citep{Booth2021AnDisk}, and extragalactic environments \citep{Aladro2013AEnvironment,Harada2021StarburstALCHEMI,Holdship2021TheALCHEMI}. In principle, where one would consider this modelling appropriate for an object, the analysis in this work allows one to plan observations to maximize information gain on the gas conditions. However, there are additional shortcomings to our model data that may limit the usefulness of this analysis even in those cases, and they require additional discussion.\par
The list of HITs is entirely dependent on our chemical modelling, which only considers static conditions. For objects such as hot cores where the conditions change on timescales much smaller than the convergence time of the HITs (see Table~\ref{table:hitlist}), the history independence will fail and successful modelling will still depend on reasonable starting abundances. Moreover, the most useful probes of the gas conditions in these objects may differ from those presented here.\par
The computed mutual information scores are more complex because they depend on both the chemical and radiative transfer models. From a chemical modelling perspective, a key issue is that we assume our gas conditions are independent when, in reality, they are not. Of particular note is the gas temperature, which is calculated as a function of the other inputs in thermochemical models. Thus, our grid covers some areas of parameter space that are not physically possible. If the grid were restricted to only include parameter sets that are achievable by a thermochemical code, then the ranking of features may change. \par
Finally, RADEX is an effective but simple radiative transfer program. The large velocity gradient method fails at high optical depth \citep{VanderTak2007}, and therefore very optically thick lines may be less informative than indicated in this work. This is of particular concern for CO, for which many transitions appear in the top feature lists for multiple physical parameters as its emission is often optically thick. However, due to how commonly CO lines are targeted, an observer is likely to know in advance whether a CO line will be optically thick and can thus exclude it from consideration when selecting transitions.\par
Finally, RADEX does not include treatment for masing, infrared pumping, or the overlap of spectral lines from different molecules. As such, if any of these effects are expected to strongly affect the intensity of a molecular line in a specific object, the intensity of that line may not be as informative about the physical conditions as our modelling would indicate.\par
\section{Conclusions}
\label{sec:conclusions}
In this work we have used chemical and radiative transfer modelling to build a large dataset of synthetic molecular transition fluxes across a wide range of physical parameters. By using different gas histories and comparing final abundances, we have identified `HITs': the species that most quickly forget their history, making them simple to model. We then used the mutual information between the line intensities of our HITs and the physical parameters to determine which transitions are most useful for constraining the gas conditions of an astrophysical system.\par
By defining a HIT as a species whose abundance varies by less than 0.5 dex after 1 MYr across a large range of physical conditions (80\% of our parameter grid), we find 43 HITs and were able to calculate the line intensities for 23 of them using RADEX. We find that our HITs typically forget their history in less than 0.1 MYr.\par
From the mutual information between the HIT transition intensities and the physical parameters, we provide a list of 19 pairs of species that have highly informative transition ratios for measuring at least one physical parameter.\par
We demonstrate the effectiveness of this method of line selection by training random forest regressors to predict the physical parameters using the most informative line ratios from our dataset.\ We show that small errors can be obtained on most physical parameters by choosing lines based on their mutual information with the target physical parameter.\par
We find that many highly ranked features have a large amount of information in common with each other. Therefore, efficient line selection would consider only lines that have low mutual information with previously selected lines as well as high mutual information with the target parameter. In future work, we will find an efficient method to filter on this metric and provide an online tool for observation planning.
\begin{acknowledgements}
We thank the referee for their comments which greatly improved this manuscript. This work is part of a project that has received funding from the European Research Council (ERC) under the European Union’s Horizon 2020 research and innovation programme MOPPEX 833460.
\end{acknowledgements}
\bibpunct{(}{)}{;}{a}{}{,} % to follow the A&A style
\bibliographystyle{aa}
\bibliography{references}
\end{document}